\documentclass[amssymb, floatfix, superscriptaddress, aps, prx, reprint]{revtex4-1}
\usepackage{amsmath} %\text{} for text in math
\usepackage{bm}  %\bm{} for bold font in math
\usepackage{graphicx}
\usepackage{physics}
\usepackage{color}
\usepackage[colorlinks=true, citecolor=blue, linkcolor=blue, urlcolor=blue]{hyperref}
\graphicspath{ {./}{./figures/} }

\begin{document}
\setlength{\intextsep}{5pt}
\title{Predicting charge transport in the presence of polarons: \\
The beyond-quasiparticle regime in SrTiO$_{\mathrm{3}}$}

\author{Jin-Jian Zhou}
\affiliation{Department of Applied Physics and Materials Science, California Institute of Technology, Pasadena, CA 91125, USA.}
\author{Marco Bernardi}
\email[E-mail: ]{bmarco@caltech.edu}
\affiliation{Department of Applied Physics and Materials Science, California Institute of Technology, Pasadena, CA 91125, USA.}
%\date{\today}
%
%
%. ABSTRACT
%
\begin{abstract} \label{abstract}
In materials with strong electron-phonon ($e$-ph) interactions, the electrons carry a phonon cloud during their motion, forming quasiparticles known as polarons. 
Predicting charge transport and its temperature dependence in the polaron regime remains an open challenge. 
Here, we present first-principles calculations of charge transport in a prototypical material with large polarons, SrTiO$_{3}$. %, in which the transport mechanisms are controversial. 
Using a cumulant diagram-resummation technique that can capture the strong $e$-ph interactions, our calculations can accurately predict the experimental electron mobility in SrTiO$_{3}$ between 150$-$300~K. 
They further reveal that for increasing temperature the charge transport mechanism transitions from band-like conduction, in which the scattering of renormalized quasiparticles is dominant, 
to a beyond-quasiparticle transport regime governed by incoherent contributions due to the interactions between the electrons and their phonon cloud. Our work reveals long-sought microscopic details of charge transport in SrTiO$_{3}$, 
and provides a broadly applicable method for predicting charge transport in materials with strong $e$-ph interactions and polarons. \\
\end{abstract}
\maketitle
%

%
% INTRODUCTION
\section{introduction}
\vspace{-10pt}
\indent
Understanding charge transport in complex materials is a grand challenge of fundamental and technological relevance. 
The interactions between electrons and phonons (the quanta of lattice vibrations) set an intrinsic limit for the conductivity and typically control charge transport near room temperature. 
% long-lived
When electron-phonon ($e$-ph) interactions are weak, charge transport is well described by the scattering of quasiparticles (QPs)~\cite{Ziman2007}, leading to the well-known band-like conduction regime.
As the $e$-ph interactions become stronger, the electrons are dressed by a cloud of phonons, forming composite charge carriers known as polarons~\cite{Frohlich1954,Holstein1959,Devreese2009}. 
In the limit of strong $e$-ph coupling, the electrons are self-trapped by the lattice distortions, and the conduction mechanism becomes the thermally activated hopping of localized polarons~\cite{emin2012}. \\
%
% LARGE POLARON
% 
\indent
Many oxides and organic crystals exhibit $e$-ph coupling strengths intermediate between the band-like and polaron hopping limits~\cite{Iadonisi2006,Fratini2016}. 
In this so-called \lq\lq large polaron'' regime, the charge transport mechanisms and their temperature dependence are not well understood.   
%one-dimensional and temperature 
The transition from band-like to hopping conduction for increasing $e$-ph coupling strength is also unclear, 
and recent work on the Holstein model uncovered an incoherent transport regime at intermediate coupling~\cite{Mishchenko2015}.
%in the intermediate $e$-ph coupling at finite temperature of this incoherent transport regime is missing, so-called large polaron However, p
Yet, predictive calculations and microscopic understanding of charge transport in the intermediate $e$-ph coupling regime remain an open challenge. \\
% present in a wide range of materials are still
\indent
%
%  STO
% 
Strontium titanate (SrTiO$_{3}$), which is stable in the cubic phase above 105~K, is a prototypical material with intermediate $e$-ph coupling in which large polaron effects are clearly seen in experiments~\cite{vanMechelen2008,Devreese2010,Chen2015,Wang2016}.
Charge transport in SrTiO$_{3}$ is a decades-old problem~\cite{Frederikse1967,Wemple1969,Verma2014,Zhou2016a,Cain2013,Lin2017}, 
yet its underlying microscopic mechanisms are still debated~\cite{Collignon2019}. The electron mobility in cubic SrTiO$_{3}$ exhibits a roughly $T^{-3}$ temperature dependence above 150~K~\cite{Cain2013,Lin2017}, which is commonly attributed to the scattering of electron QPs with phonons~\cite{Frederikse1967,Wemple1969,Verma2014,Zhou2016a}. 
%, even though details such as which phonons contribute the most to scattering have remained controversial~\cite{Frederikse1967,Wemple1969,Verma2014,Zhou2016a}.  
Different phenomenological models based on QP scattering have been proposed that can fit the experimental transport data~\cite{Verma2014,Mikheev2015}.
However, the carrier mean free paths extracted from experiment in SrTiO$_{3}$ fall below the interatomic distance~\cite{Lin2017}, violating the Mott-loffe-Regel (MIR) criterion for the applicability of the QP scattering picture~\cite{Hussey2004}. 
While there is consensus that large polarons are present in SrTiO$_{3}$~\cite{vanMechelen2008,Devreese2010,Chen2015,Wang2016,Devreese2016}, charge transport in this regime cannot yet be predicted, and 
detailed microscopic understanding has remained elusive~\cite{Collignon2019}.\\
%
%the microscopic mechanisms of charge transport in cubic SrTiO$_{3}$ are still unclear and debated.
%
%
% AB INITIO
%
\indent
First-principles calculations based on lowest-order $e$-ph scattering plus the Boltzmann transport equation (BTE)~\cite{Bernardi2016} can accurately predict the conductivity in simple metals and semiconductors~\cite{Mustafa2016,Zhou2016,Liu2017,Ma2018,Lee2018}. %and related oxides exhibiting large polarons a significant discrepancy emerges between the BTE approach and experiments. 
%In SrTiO$_{3}$, even when scattering with all phonon modes (included the soft modes) is correctly taken into account,
We have recently shown~\cite{Zhou2018} that when this approach is applied to SrTiO$_{3}$, one can obtain an accurate temperature dependence for the electron mobility if all the phonons 
(including the soft modes) are taken into account. However, the absolute value of the computed electron mobility is an order of magnitude greater than experiment~\cite{Zhou2018}. It is clear that QP scattering alone cannot explain charge transport in SrTiO$_{3}$, consistent with the MIR limit violation; a fully quantum mechanical framework is needed to predict charge transport in this beyond-QP regime. \\
%
% HERE  WE  SHOW
%
\indent
Here we show first-principles calculations of charge transport in cubic SrTiO$_{3}$ using a finite-temperature retarded cumulant approach that includes higher-order $e$-ph interactions and goes beyond the QP scattering picture. 
%combine them with the Kubo formula to compute the electron mobility in a formalism that goes 
%
%using the Kubo formula that goes beyond the BTE approach and 
%using a finite-temperature retarded cumulant approach combined with the Kubo formula. 
%Crucial to our method is its ability to include higher-order $e$-ph Feynman diagrams. 
Our calculations can accurately predict the experimental electron mobility in SrTiO$_{3}$ between 150$-$300~K, and further shed light on its microscopic origin. 
%via the analysis of the electron spectral function and low energy optical conductivity.
We show that the weight of the QP peak in the electron spectral function is strongly renormalized, with significant weight transfer to the incoherent phonon satellites. %the QP peak even disappears at momenta beyond a certain value.
%, leaving the spectral function fully incoherent. 
While the renormalized QPs control transport at low temperature, the incoherent contributions from the phonon satellites and large momentum states %at momenta at which the QP peak has disappeared, 
are significant at room temperature, indicating a transport regime beyond the QP scattering paradigm. Consistent with these trends, our analysis shows that near room temperature the scattering rate (extracted from the optical conductivity) breaks the Planckian limit of $k_{\rm B}T$ for semiclassical transport~\cite{Hartnoll2015}, clearly indicating a beyond-QP charge transport regime. 
%
% the incoherent phonon satellites and momentums without QP contributes significantly to transport as $T$ approaching room temperature, indicating a transport regime beyond the QP scattering-based paradigm.
Our work opens new avenues for computing charge transport in complex materials with large polarons and beyond the QP scattering regime.  \\
%beyond the QP scattering regime in complex materials with large polarons.
% in transition metal oxides, organic crystals and other
%
\begin{figure}%[!htbp]
\includegraphics[width=\columnwidth]{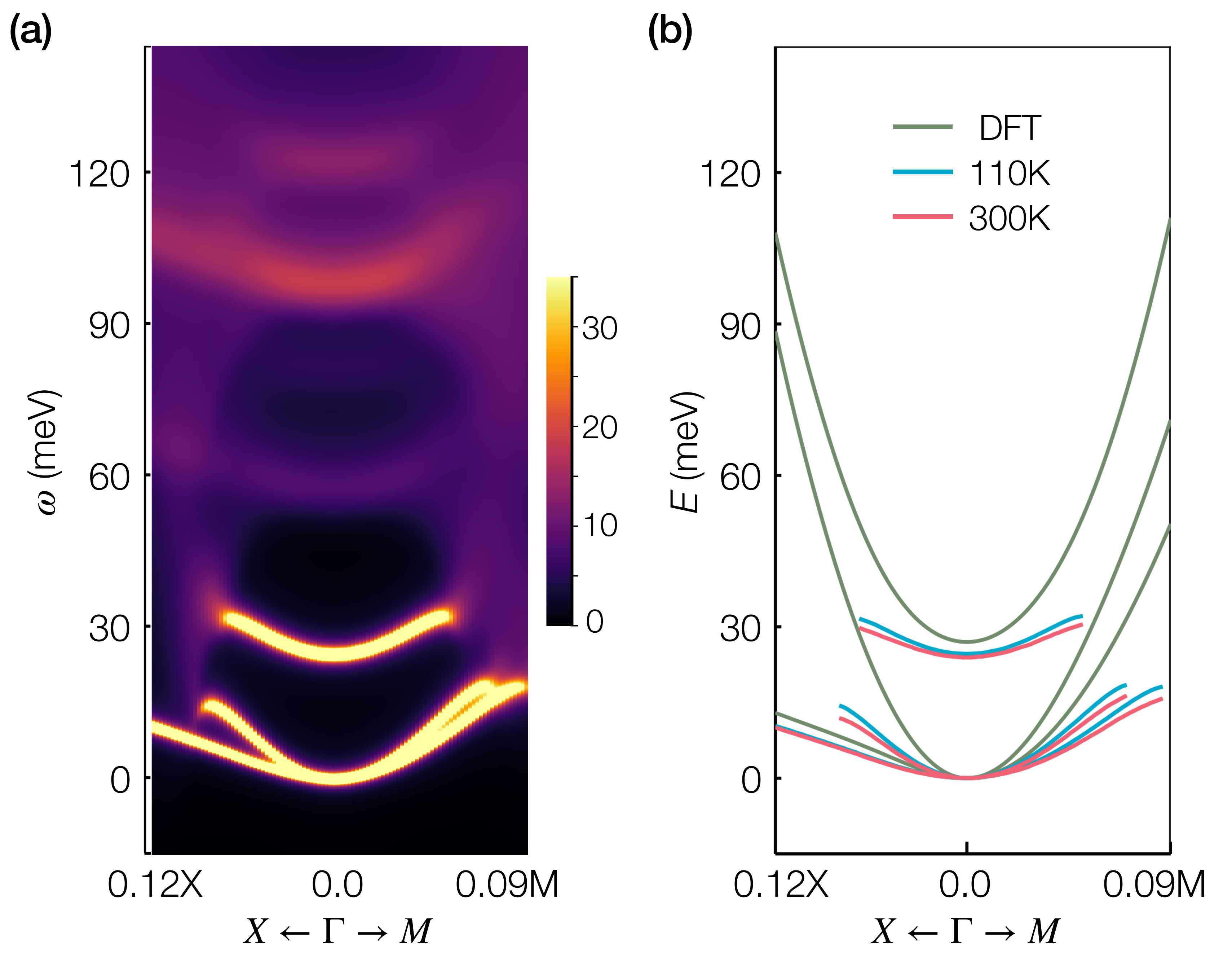}
\caption{ (a) Combined spectral functions $A_{n\bm{k}}(\omega)$ for the three lowest conduction bands in cubic SrTiO$_{3}$, for $\bm{k}$ along the $\Gamma$$-$$X$ and $\Gamma$$-$$M$ Brillouin zone directions at $110$~K. (b) The energy-momentum dispersion of the QP peaks of the spectral function, shown at $110$~K and $300$~K, compared with the electronic band structure from DFT. The zero of the energy axis is set to the conduction band minimum.
} \label{fig:qpdisp}
\end{figure}
%
%  ELECTRON SPECTRAL FUNCTION
% 
%
\section{Electron spectral function}
\vspace{-10pt}
\indent
Central to our approach for computing charge transport is the electron spectral function, $A(\omega)$, which can be seen as the density of states of a single electron. 
Due to the interactions with the phonons, the spectral function consists of a QP peak representing a single-electron-like excitation  
and an incoherent part including both phonon satellite peaks and a background contribution~\cite{Damascelli2003,Martin2016}. 
While the spectral weights of the QP peak and incoherent part may vary, they always
add up to one due to the sum rule $\int\!d\omega A(\omega) = 1$, which amounts to a conservation of the electron.
To investigate the dynamics of the electrons and their interactions with the phonons, we compute the spectral function 
with a finite-temperature retarded cumulant approach that can account for higher-order $e$-ph interactions (see Appendix~\ref{app:cum}), and use it directly to predict transport, without relying on QP scattering approaches.\\
%
% METHODS (BRIEFLY)
%
\indent
% EL, PH, E-PH
We focus on the electron spectral function in cubic SrTiO$_{3}$ above 110~K. 
Leveraging our recently developed approach~\cite{Zhou2018}, which combines density functional theory (DFT), its linear response extension~\cite{Baroni2001} and the temperature dependent effective potential (TDEP) method~\cite{Hellman2011}, we compute the band structure, lattice vibrations and $e$-ph interactions for all phonon modes, including the soft modes due to the lattice anharmonicity (see Appendix~\ref{app:detail}).
% CUMULANT
Using these quantities, we calculate the spectral function using a 
retarded cumulant formalism~\cite{Kas2014,Kas2017}. 
The retarded cumulant approach is based on an exponential ansatz for the retarded Green's function of an electronic state $\left|n\bm{k}\right\rangle$ (with band index $n$ and momentum $\bm{k}$) in the time domain:
% CUMULANT
\begin{equation}
G^{R}_{n\bm{k}}(t) = -i\theta(t) e^{-i\varepsilon_{n\bm{k}} t}e^{C_{n\bm{k}}\left(t\right)} \,\,,
\label{eq:cum_exp}
\end{equation}
where $t$ is time, $\theta(t)$ is the Heaviside step function, and $\varepsilon_{n\bm{k}}$ is the non-interacting electron energy from DFT. The $e$-ph interactions are included in the cumulant $C_{n\bm{k}}\left(t\right)$; 
an approximate expression derived in Ref.~\cite{Kas2014} computes $C_{n\bm{k}}\left(t\right)$ using the off-shell lowest-order $e$-ph self-energy, $\Sigma_{n\bm{k}}(\omega)$~\cite{Bernardi2016} as input:
\begin{equation}
 C_{n\bm{k}}\left(t\right) = \int_{-\infty}^{\infty}d \omega \frac{\beta_{n\bm{k}}  \left(\omega \right)}{\omega^{2}}\left(e^{-i\omega t}+i\omega t-1\right)\,\,,
 \label{eq:cumulant}
\end{equation}
where $\beta_{n\bm{k}}(\omega) \equiv |\mathrm{Im}\Sigma_{n\bm{k}} \left(\omega +\varepsilon_{n\bm{k}}\right)|/\pi$. 
The spectral function can be obtained for each state from the retarded Green's function in the frequency domain, using $A_{n\bm{k}}(\omega) = -\mathrm{Im}G_{n\bm{k}}^{R}(\omega)/\pi$. 
%
%
%finite-temperature retarded cumulant method developed in this work (see Methods). 
%In this framework, the $e$-ph interactions for each electronic state $\left|n\bm{k}\right\rangle$ with band index $n$ and momentum $\bm{k}$ are included 
%in the one-particle retarded Green's function $G_{n\bm{k}}^{R}$ via the so-called cumulant function $C_{n\bm{k}}\left(t\right)$: 
%
%\begin{equation}
%G^{R}_{n\bm{k}}(\omega) = -i \int^{\infty}_{0} e^{i(\omega-\varepsilon_{n\bm{k}}) t}e^{C_{n\bm{k}}\left(t\right)} dt\,\, ,
%\label{eq:cum_exp}
%\end{equation}
%where $\varepsilon_{n\bm{k}}$ is the DFT band electron energy and $C_{n\bm{k}}\left(t\right)$ is obtained from the off-shell lowest-order $e$-ph self-energy, $\Sigma_{n\bm{k}}(\omega)$~\cite{Bernardi2016}, as
%\begin{equation}
 %C_{n\bm{k}}\left(t\right) = \int_{-\infty}^{\infty}d \omega \frac{|\mathrm{Im}\Sigma_{n\bm{k}} \left(\omega +\varepsilon_{n\bm{k}}\right)|}{\pi \omega^{2}}\left(e^{-i\omega t}+i\omega t-1\right).
 %\label{eq:cumulant}
%\end{equation}
%
%The spectral function is then obtained from the retarded Green's function, using $A_{n\bm{k}}(\omega) = -\mathrm{Im}G_{n\bm{k}}^{R}(\omega)/\pi$. 
%implementation of the
Our scheme to compute the retarded Green's function at finite-temperature (see Appendix~\ref{app:cum}) further allows us to compute the spectral function as a function of temperature. 
%Our finite-temperature retarded cumulant approach (see Appendix~\ref{app:cum}) further allows us to compute the spectral function at various temperatures. 
The retarded cumulant approach includes higher-order $e$-ph Feynman diagrams beyond the Migdal approximation~\cite{Hedin1999}, and it produces accurate spectral functions (see below) that can capture the strong $e$-ph interactions. On the other hand, we have verified that the lowest-order Dyson-Migdal approximation generates spectral functions with large errors in the QP spectral weight and satellites energies, consistent with recent results at zero temperature~\cite{Nery2018}. \\
%
%that includes higher-order $e$-ph Feynman diagrams beyond the Migdal approximation~\cite{Hedin1999} could produce accurate spectral functions that can capture the strong $e$-ph interactions
%The cumulant approach includes higher-order $e$-ph Feynman diagrams beyond the Migdal approximation~\cite{Hedin1999},  thus producing accurate spectral functions that can capture the strong $e$-ph interactions, and we have verified that the Dyson-Migdal approach generates spectral function with significant errors in the QP spectral weight and satellites structures, 
%the significant errors in the spectral weight and satellites structure of the spectral function from the Dyson-Migdal approach %
%builds in implicit dynamic vertex corrections, includes higher order $e$-ph diagrams beyond the Migdal approximation, 
%with all phonon modes treated on the equal footing (see Methods), 
%Our finite-temperature implementation of the retarded cumulant approach (see Methods) further allows us to compute the spectral function at different temperatures. \\
% ~\cite{Kas2017}
%
% FIGURE 2
%
\begin{figure*}[!htbp]
\includegraphics[width=2.0\columnwidth]{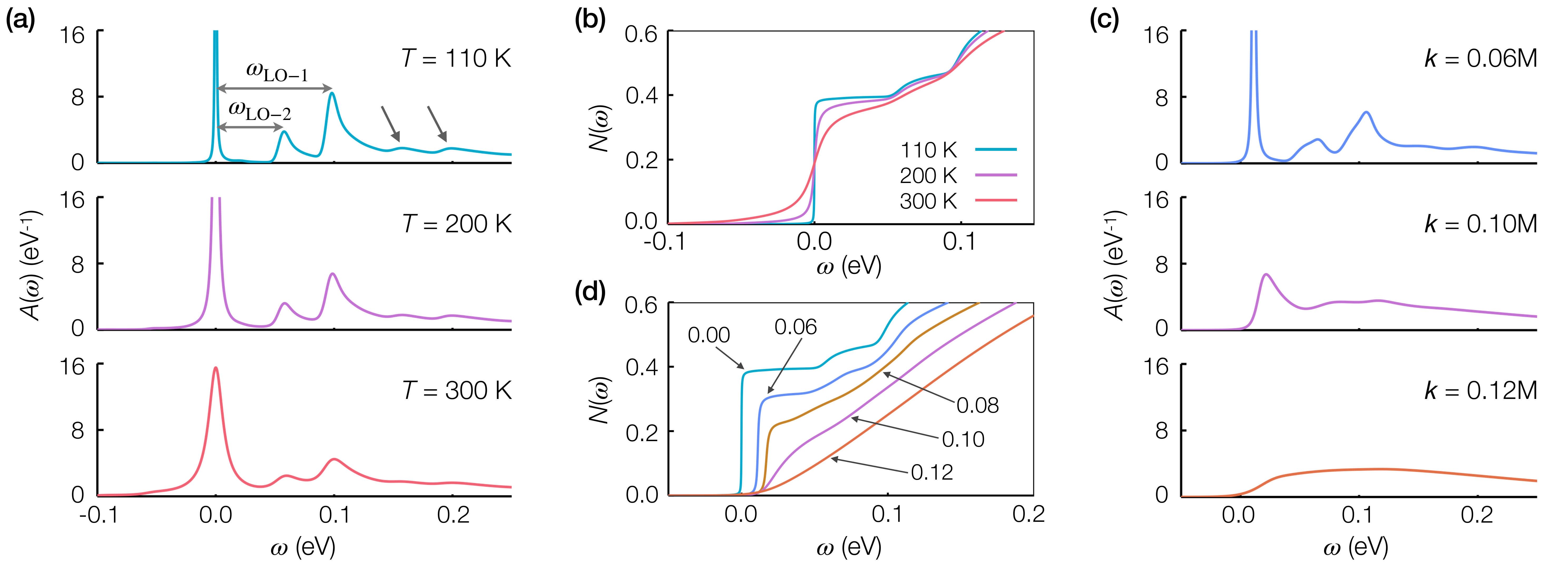}
\caption{ Computed electron spectral functions $A_{\bm{k}}(\omega)$ for the lowest conduction band in cubic SrTiO$_{3}$. In each panel, the zero of the $\omega$ energy axis is set to the energy of the QP peak at the $\Gamma$ point. 
(a) The spectral function $A_{\bm{k}}(\omega)$ at three temperatures between 110$-$300~K, for $\bm{k} = \Gamma$. The energies of the two LO phonons that couple strongly with the electrons are labelled $\omega_{\text{LO-1}}$ and $\omega_{\text{LO-2}}$, and the arrows point to the second set of satellites at $2\omega_{\text{LO-1}}$ and $\omega_{\text{LO-1}}+\omega_{\text{LO-2}}$. 
% , computed as $N(\omega) = \int_{-\infty}^{\omega} A(\omega^{\prime})d\omega^{\prime}$
(b) The spectral weight $N(\omega)$ obtained by integrating the spectral function up to an energy $\omega$. 
(c) The spectral function $A_{\bm{k}}(\omega)$ at 110~K for different values of $\bm{k}$ along the $\Gamma$$-$$M$ Brillouin zone line, and (d) The corresponding integrated spectral weight $N(\omega)$. %for $\bm{k}$ from $\Gamma$~(0.00) to 0.12$M$. %T everywhere? 
%; the temperature is kept fixed at 110~K
%spectral function for the lowest conduction band for $\bm{k}$ of 0.06$M$, 0.10$M$ and 0.12$M$, the corresponding integrated spectral weights are shown in (d). 
} \label{fig:specfunc}
\end{figure*}
%
% FIG.1a, MASS RENORMALIZATION
%
\indent
\hspace{-1pt}The computed electron spectral functions for the three lowest conduction bands in cubic SrTiO$_{3}$ at 110~K are combined in a color map and shown in Fig.~\ref{fig:qpdisp}(a). 
Each state exhibits a rather sharp QP peak at low energy and broader phonon satellite peaks at higher energies ($\sim$60 meV or more above the QP peak). 
By tracking the low-energy QP peaks, we map the energy-momentum dispersion of the QPs. Figure~\ref{fig:qpdisp}(b) shows that the interacting QPs exhibit a heavier effective mass than in the DFT band structure calculations, in which the $e$-ph interactions are not included. The mass enhancement is a factor of 1.8$-$2.6 for different bands and directions, and it increases only slightly with temperature. 
Taking the lowest bands along $\Gamma-$M as an example, the DFT effective mass is roughly 0.75$m_{e}$ ($m_e$ is the electron mass), 
compared to a QP effective mass of 1.4$m_{e}$ at 110~K and a slightly heavier mass of 1.6$m_{e}$ at 300~K. 
The mass enhancement is thus roughly a factor of 2, in excellent agreement with experimental results at light doping~\cite{vanMechelen2008,Wang2016}.  \\
%
% WEIGHT RENORMALIZATION
%
\indent
The interactions with the surrounding phonons not only make the electron QPs heavier, 
they also significantly reduce the QP spectral weight to a value well below one, 
transferring weight to the higher-energy incoherent phonon satellites. 
%of much less than unity, 
% 
The QP peak even disappears at large electron momenta, leaving a spectral function made up entirely by the incoherent background. 
These trends are analyzed in detail below and in Fig.~\ref{fig:specfunc}, focusing on how the spectral function of the lowest conduction band changes as a function of temperature and momentum. \\
% 
% SATELLITES
%
\indent
The spectral function at the conduction band minimum at $\Gamma$ exhibits a main QP peak, two main satellites (also known as phonon sidebands or replicas), and weaker additional satellites at higher energy [see Fig.~\ref{fig:specfunc}(a)]. 
The two main satellite peaks are at an energy $\omega_{\text{LO-1}}=98$~meV and $\omega_{\text{LO-2}}=57$~meV above the main QP peak; these values correspond, respectively, to the energies of the two longitudinal optical (LO) modes with long-range $e$-ph interactions that exhibit the strongest coupling with electrons~\cite{Zhou2018}.  
% HIGHER HARMONICS
Note also the presence in Fig.~\ref{fig:specfunc}(a) of weak phonon sideband peaks at energies of $2\omega_{\text{LO-1}}$ and $\omega_{\text{LO-1}}+\omega_{\text{LO-2}}$. These higher-order replicas, which are known to occur in the strong coupling limit of the Holstein model~\cite{Berciu-PRL}, are akin to the higher harmonics observed in phonon Floquet states~\cite{PhFloquet}; they are a signature of strong coupling with the LO modes.\\
%
%
%
%  FIGURE 3
%
\begin{figure*}[!t]
\includegraphics[width=1.96\columnwidth]{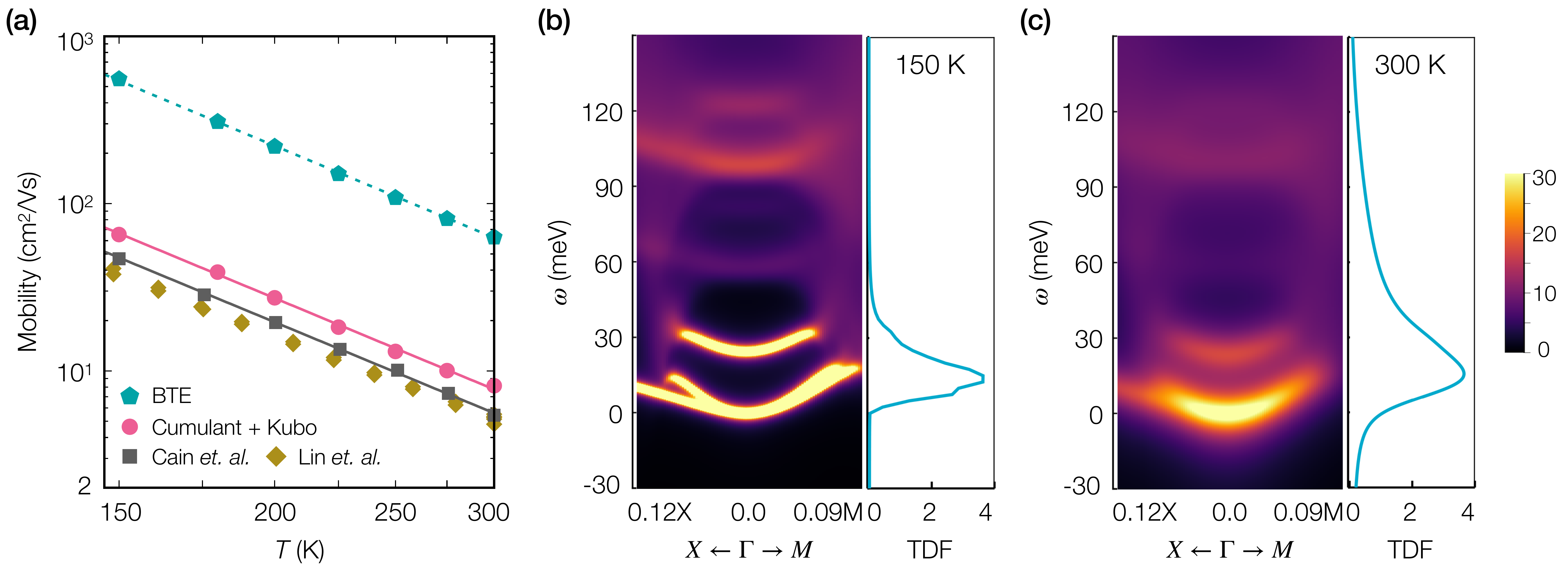}
\caption{ (a) Electron mobility as a function of temperature, computed using the retarded cumulant approach plus the Kubo formula (red circles) and compared with experimental values taken from Refs.~\cite{Cain2013,Lin2017}. 
The mobility computed in Ref.~\cite{Zhou2018} using lowest-order $e$-ph scattering plus the BTE is also shown (blue pentagons). 
(b) The combined conduction-band spectral functions at $T=150$~K are shown together with the TDF defined in Eq.~(\ref{eq:tdf}), which quantifies the contribution to the DC conductivity as a function of electron energy $\omega$. 
(c) The same quantities as in (b) shown at $T=300$~K.  
The zero of the energy axis is set to the energy of the lowest QP peak.
} \label{fig:mobility}
\end{figure*}
\indent
\hspace{-1pt}The main phonon sidebands are associated with the polaron plus one-phonon continuum, and are a hallmark of the large polaron regime~\cite{Mahan2000}.
Note that our calculations are performed on lightly $n$-doped SrTiO$_{3}$, with the chemical potential lying below the lowest QP peak; therefore, 
as expected, the phonon satellites appear at energy \textit{higher} than the QP peak as they correspond to the excitation of an   ``electron-like'' QP plus one LO phonon~\cite{Damascelli2003,Nery2018}.
On the other hand, recent angle-resolved photoemission measurements on heavily $n$-doped samples, in which the chemical potential is above the QP peak, revealed a phonon sideband roughly $\sim$100~meV \textit{below} the QP peak, which corresponds to the excitation of a ``hole-like'' QP plus one LO phonon~\cite{Chen2015,Wang2016, Damascelli2003}. 
The energy difference between the QP peak and phonon sideband observed in experiments is in very good agreement with the $\omega_{\text{LO-1}}=98$~meV energy difference between the QP peak and the most intense satellite peak found in our computed spectral function. We have verified that at high doping our approach also gives satellites with energies lower than the QP peak, consistent with experiment. The fact that the satellite position can be higher or lower than the QP peak depending on doping is well known~\cite{Damascelli2003,Zhou2018b}, but since inverse photoemission measurements are challenging, the low doping regime we compute here is rarely probed experimentally.\\ 
%in photoemission experiments.
%
%
%in the phonon cloud, which are analogous to the cavity modes in Floquet theory.
\indent
To compute the spectral weights of the QP peak and incoherent part, we integrate the spectral function up to an energy $\omega$, obtaining the spectral weight 
$N(\omega) = \int_{-\infty}^{\omega} A(\omega^{\prime})d\omega^{\prime}$ given in Fig.~\ref{fig:specfunc}(b).
%we integrate the spectral function up to an energy $\omega$, computed as $N(\omega) = \int_{-\infty}^{\omega} A(\omega^{\prime})d\omega^{\prime}$, to show the spectral weight of the coherent QP (Fig.~\ref{fig:specfunc}b). 
%, corresponding to energies of the two longitudinal optical (LO) phonons that have the strongest coupling with electrons~\cite{Zhou2018}. 
%This phonon satellite structure is a hallmark of large polaron.
We find that the spectral weight of the QP peak is $\sim$0.4 at 110~K, and thus much less than the unit value of the weak $e$-ph interaction limit. 
As the temperature increases from 110~K to 300~K, both the QP peak and the phonon satellites are broadened and smeared out [see Fig.~\ref{fig:specfunc}(a)], 
but the QP spectral weight changes only slightly, primarily due to an overlap between the QP peak and the phonon satellites near 300~K. 
%We conclude that 
At all temperatures between 110$-$300~K, the QP weight is strongly renormalized to a value of $\sim$0.4, 
implying that (pictorially) only half of the electron resides in the QP state, while the other half contributes to the incoherent dynamical interactions with the phonons.\\
% cloud.\\
%
% QP PEAK END POINT 
%
\indent
Figure ~\ref{fig:specfunc}(c) reveals the disappearance of the QP peak at large enough momentum $\bm{k}$ by showing how the spectral function changes as we increase $\bm{k}$ along the $\Gamma$$-$$M$ Brillouin zone line. 
We find that the QP spectral weight decreases with increasing momentum~[see Fig.~\ref{fig:specfunc}(d)] and that the QP peak ultimately disappears at $\bm{k}=0.12$$M$, leading to a fully incoherent spectral function at larger momenta. 
% %as the momentum $\bm{k}$ increases and moves towards the $M$ point, the QP peak shifts upward due to increasing kinetic energy and starts to mix the the phonon continuum (Fig.~\ref{fig:specfunc}c).
These so-called end points of the QP peak, which have been predicted in both the Fr{\"o}hlich~\cite{Mishchenko2000} and Holstein models~\cite{Goodvin2010}, 
are yet another signature of the strong $e$-ph interactions.
%surprising consequence  
%Their origin can be attributed to a superposition between a band-like QP and a phonon state evolving from polaron-like at small momentum to phonon-like near the end point, 
%where the phonon carries nearly all of the momentum~\cite{Mishchenko2000}. 
%
The decrease of the QP spectral weight and the disappearance of the QP peak at large momentum have a significant impact on transport, as we discuss below.
%ELSEWHERE: Such beyond-QP effects have been overlooked by theories based on the QP scattering paradigm.\\
%
%
\section{Electron mobility}
\vspace{-10pt}
\indent
% PREAMBLE: INTUITION ON LARGE POLARONS
Large polaron transport is commonly believed to be the band-like conduction of QPs with enhanced effective mass.  
However, this simplified picture neglects the fact that the QP weight can drop to values much smaller than one (here, to roughly 0.4, as discussed above), and that the contribution to transport from the incoherent part of the spectral function can be significant. 
Temperature also plays a primary role. At low temperature, the electrons occupy the low-energy QP states, and there are only few LO phonons due to their relatively high energy.
%there are fewer LO phonons due to the relatively high LO phonon energy. 
%the LO phonon cloud is small due to the relatively high LO phonon energy. 
%thermal fluctuations excite the electrons outside of the QP peak into the higher-energy window governed by incoherent contributions.
As the temperature increases, thermal fluctuations push the electrons to higher energies, exciting the electrons outside of the QP peak into the incoherent regime. 
%the tail of the electron distribution extends into the higher-energy window governed by incoherent contributions.
In addition, the number of LO phonons grows rapidly with temperature, leading to strong dynamical interactions between the electron and its phonon cloud. The incoherent contributions are thus expected to significantly influence transport at higher temperatures. \\ %HIGH ENOUGH
%
% KUBO
%
\indent
To investigate these points quantitatively in SrTiO$_3$, we compute the conductivity directly from the spectral functions $-$ therefore including both the QP and incoherent contributions $-$ using the Kubo formula~\cite{Mahan2000}. In the absence of current-vertex corrections, the conductivity can be expressed as~\cite{Economou2006,Basov2011}
\begin{equation}
\begin{split}
\sigma_{\alpha\beta}(\omega)   = & \frac{\pi\hbar e^{2}}{V_{\text{uc}}}\int d\omega^{\prime}
\frac{f(\omega^{\prime})-f(\omega^{\prime}+\omega)}{\omega}  \\
 & \times \sum_{n\bm{k}}v_{n\bm{k}}^{\alpha}\,v_{n\bm{k}}^{\beta}\,A_{n\bm{k}}(\omega^{\prime})\,A_{n\bm{k}}(\omega^{\prime}+\omega)\,\,,
\end{split}
\label{eq:opcond}
\end{equation}
where $\bm{v}_{n\bm{k}}$ is the band velocity of the electronic state $\left|n\bm{k}\right\rangle$, $f(\omega)$ is the Fermi-Dirac distribution, $V_{\text{uc}}$ is the unit cell volume, and $\alpha$ and $\beta$ are Cartesian directions. 
%Eq.~\ref{eq:opcond} has been used recently to successfully capture electron-electron interactions in strongly correlated materials~\cite{Basov2011}.
We also compute the DC conductivity, using $\sigma^{dc} = \sigma(\omega\rightarrow0)$, and the electron mobility as $\mu = \sigma^{dc} / n_{c}e$, where the carrier concentration $n_{c}$ is computed as 
\begin{equation}
n_{c}=\sum_{n\bm{k}} \int_{-\infty}^{\infty}{d\omega}f(\omega)A_{n\bm{k}}(\omega).
\label{eq:nc}
\end{equation}
We perform mobility calculations in the lightly doped regime, with electron concentrations ranging between $10^{17}$$-$$10^{18}$ cm$^{-3}$, 
and find that the computed mobility is nearly independent of the chosen concentration above 150~K, consistent with experimental data~\cite{Cain2013,Lin2017}.\\
%
%
% FIGURE 4
%
\begin{figure}[!b]
\includegraphics[width=\columnwidth]{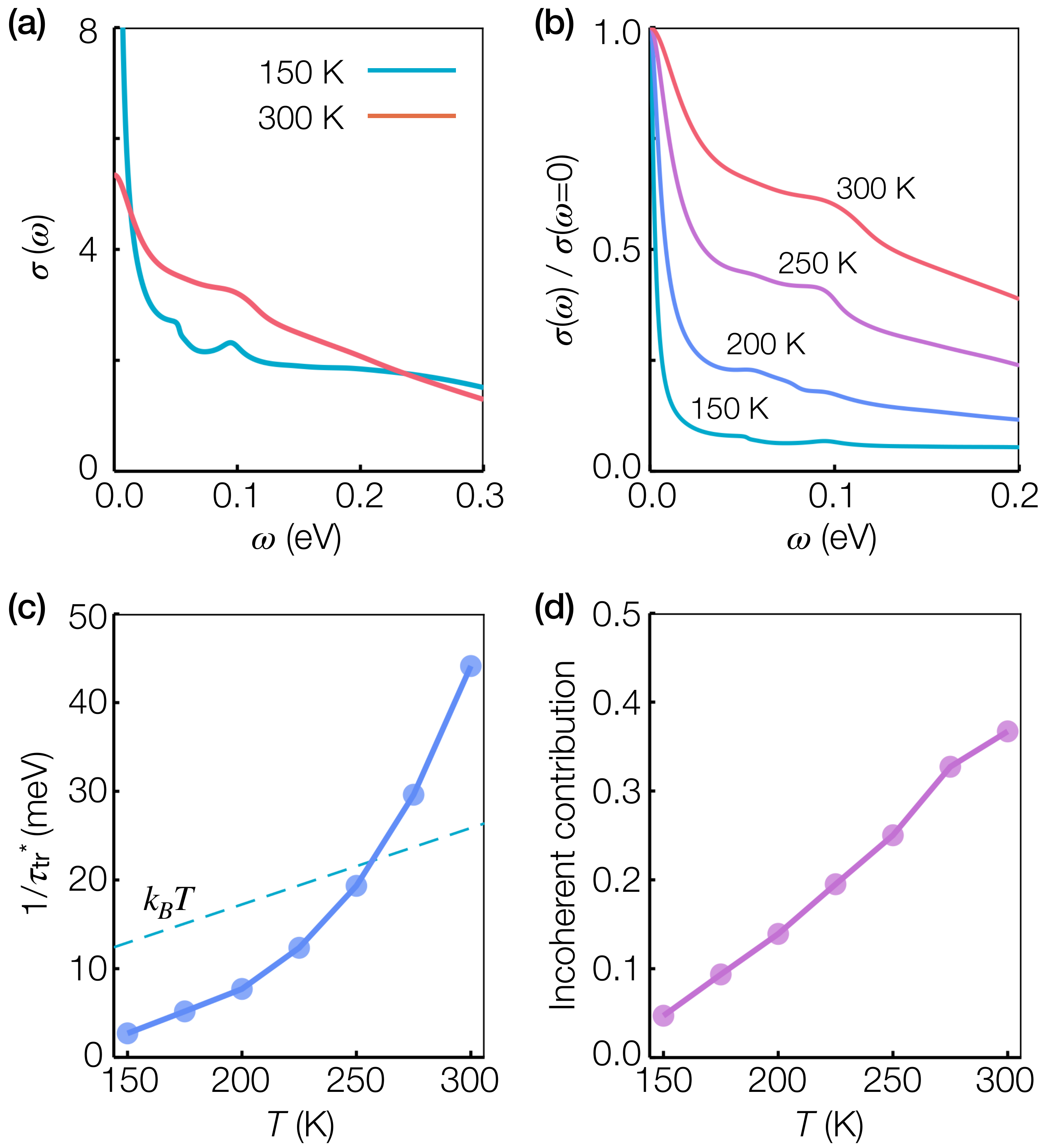}
\caption{ (a) Comparison between the computed optical conductivities at 150~K and 300~K. The curves were normalized to possess the same integral. 
%To make the comparison meaningful, the optical conductivity is normalized by its integral at each temperature.
(b) Computed optical conductivity divided by the DC conductivity at each temperature. 
%normalized to the DC conductivity. 
(c) The inverse of the effective transport relaxation time, $\tau_{\text{tr}}^{*\,-1}\,$, extracted from $\sigma(\omega)$ and shown as a function of temperature. The Planckian limit $k_{\rm B}T$ is shown with a dashed line. 
(d) The incoherent contribution to charge transport, quantified by the DC conductivity ratio $\sigma^{dc}_{\rm inc} / \sigma^{dc}$ defined in the text. 
%where the incoherent conductivity $\sigma_{\rm inc}$ is obtained by integrating the TDF at energies greater than all QP peaks. 
} \label{fig:optical} %$\omega$ > 32~meV [see Fig.~\ref{fig:mobility}(b,c)]
\end{figure}
\indent
Our computed electron mobility in SrTiO$_{3}$ as a function of temperature is shown in Fig.~\ref{fig:mobility}(a) and compared with experimental data taken from Refs.~\cite{Cain2013,Lin2017}.
Both the absolute value and the temperature dependence of the computed mobility are in excellent agreement with experiments. Our computed mobility at 300~K is about 8 cm$^{2}$/V$\,$s, 
versus an experimental value of $\sim$5 cm$^{2}$/V$\,$s~\cite{Cain2013,Lin2017}. %LEAVE OUT: Jalan2018,
% BTE vs. CUMULANT
For comparison, the mobility obtained using lowest-order $e$-ph scattering plus the BTE is an order of magnitude higher than experiments~\cite{Zhou2018}, as also shown in Fig.~\ref{fig:mobility}(a). \\
\indent
The order-of-magnitude mobility drop from the BTE to the cumulant approach in SrTiO$_{3}$ is due to both the QP renormalization and the incoherent contributions. 
The cumulant approach clearly shows that the QP peak is strongly renormalized with significant weight transfer to higher energy incoherent satellites, and the QP peak even disappears as momentum $\bm{k}$ goes beyond the end points [see Fig.~\ref{fig:specfunc} and Fig.~\ref{fig:mobility}(b,c)]. 
However, the BTE inherently assumes that the QP states are well defined for all bands and momentum values, so it fails to capture the contributions from the incoherent part of the spectral function, thus placing all the weight in the QP peak and significantly overestimating the mobility. 
%the loss of spectral weight in the QP peak 
Our results show unambiguously that the established lowest-order $e$-ph plus BTE 
approach~\cite{Bernardi2016,Mustafa2016,Zhou2016,Liu2017,Ma2018,Lee2018} is accurate only in the case of weak $e$-ph interactions. 
Including higher-order $e$-ph processes and incoherent polaron effects via the cumulant approach greatly improves the computed mobility in materials with strong $e$-ph coupling. \\
%
%
% ROLE OF SOFT MODES
%Note that both the BTE and the cumulant calculations give an accurate temperature dependence for the mobility, a consequence of the fact that both approaches rely on accurate phonon dispersions 
%and treat $e$-ph coupling to all the phonon modes on the same footing, including the important soft modes~\cite{Zhou2018}. 
%If the ferroelectric soft modes are neglected in the calculation, one would get a significant error in the mobility, especially at low temperature, 
%and both approaches would give an inaccurate temperature dependence of the mobility, as shown in Supplementary Fig.~S1[SM]. 
%.
\indent
Correctly taking into account the contributions from all the phonon modes, including the soft modes~\cite{Zhou2018}, is also essential for predicting the electron mobility and its temperature dependence in SrTiO$_{3}$. Neglecting the soft mode in the calculations leads to significant errors in the computed mobility and its temperature dependence~(see Fig.~\ref{fig:soft} in Appendix~\ref{app:softmode}), both in the BTE and in the cumulant approach. 
These results show that the soft modes, and not just the LO phonons as is widely believed, also contribute to charge transport in the large polaron regime. \\
\indent
Lastly, note also that in the weak $e$-ph coupling limit, in which the spectral function consists only of a sharp QP peak and no incoherent contributions, the expression we use for $\sigma^{dc}$ reduces to the conductivity obtained from the relaxation time approximation (RTA) of the BTE~\cite{Economou2006,Allen2015}. 
For a material with weak $e$-ph interactions and negligible polaron effects, one thus expects that the cumulant and BTE approaches predict the same transport results. We perform this sanity check for GaAs, showing that its mobility curves computed with the cumulant and BTE-RTA are in close agreement (see Fig.~\ref{fig:gaas} in Appendix~\ref{app:gaas}).
In the case of SrTiO$_{3}$, one the other hand, it is apparent that the BTE cannot predict the mobility correctly. \\
%
% COHERENT VS INCOHERENT CONTRIBUTIONS
% 
%
\section{Coherent and incoherent contributions}
\vspace{-10pt}
\indent
\hspace{-1pt}Our approach for computing the conductivity in the presence of polarons further allows us to resolve the coherent and incoherent contributions to transport and to uncover different transport regimes as a function of temperature. %analyze
We write the DC conductivity as $\sigma^{dc} =  \int \Phi(\omega) d\omega$, where the integrand
\begin{equation}
\Phi(\omega) = \frac{\pi\hbar e^{2}}{V_{\text{uc}}}\sum_{n\bm{k}} v_{n\bm{k}}^{\alpha}\,v_{n\bm{k}}^{\beta} \left|A(n\bm{k}, \omega)\right|^{2} \left(-\frac{\partial f(\omega)}{\partial \omega} \right),
\label{eq:tdf}
\end{equation}
is referred to as the transport distribution function (TDF), and is employed here to quantify the contributions to the DC conductivity as a function of electron energy $\omega$. 
% TDF IN FIG 3b,c
We analyze the TDF in the 150$-$300 K range in Fig.~\ref{fig:mobility}(b,c)$\,$, and find that both the coherent QP peak and the incoherent part contribute to transport. At temperatures below $\sim$200~K, the TDF spans primarily the QP peaks [see Fig.~\ref{fig:mobility}(b)], implying that transport is dominated by the scattering of QPs with a strongly renormalized spectral weight. 
%$-$ the energy range contributing to the conductivity, hence transport can be described by the BTE and the scattering of QPs as long as the significant QP weight renormalization is taken into account.
%
As the temperature increases, the high-energy tail of the TDF extends over the incoherent contributions [see Fig.~\ref{fig:mobility}(c)], which become important and contribute by as much as 40\% to the conductivity at 300 K, as discussed below.
Therefore, transport near room temperature is governed not only by the weight-renormalized QPs, but also by the incoherent phonon satellites above the QP peaks and by the polaron states at large momenta beyond the end points, where the QP peaks disappear. 
The picture of QP scattering is inadequate to describe transport at room temperature in SrTiO$_3$, and a more complex picture emerges in which transport is an interplay between the QP renormalization and the contributions from the incoherent phonon sidebands and from the polaron states beyond the end points, all of which are consequences of the dynamical interactions between the electrons and the phonon cloud.\\
%
%a delicate interplay between the QP renormalization,  
%mass enhancement and loss of spectral weight of the QPs, combined with 
%the contributions from the incoherent phonon sidebands, and the polaron states beyond the end points, 
%\indent
%
%
%. OPTICAL CONDUCTIVITY
%
\indent
These conclusions are supported by experiments on the optical conductivity. %which we can explain and successfully model. 
Recent experiments in SrTiO$_{3}$ show that the Drude peak at low frequency in the optical conductivity, which is associated with the coherent band-like transport of QPs, loses weight for increasing temperatures up to 300 K~\cite{vanMechelen2008,Devreese2010}. 
Figure~\ref{fig:optical}(a) compares the low-energy optical conductivities at 150~K and 300~K,   
both computed with Eq.~(\ref{eq:opcond}) and normalized to possess the same integral, consistent with the optical sum rule.
The optical conductivities exhibit a Drude-like peak centered at zero frequency, and an incoherent shoulder structure consisting of phonon sidebands plus a broad background. We find a significant weight transfer from the Drude peak to the incoherent shoulder as the temperature increases from 150~K to 300~K, in agreement with experiments~\cite{vanMechelen2008}.
The Drude peak is sharp at 150~K, but it broadens rapidly as the temperature increases [see Fig.~\ref{fig:optical}(b)].
These trends confirm the transition seen in our transport results from a renormalized QP regime at low temperature to an incoherent, beyond-QP regime near room temperature.\\ 
%
% 
%Figure~\ref{fig:optical}a shows the low-energy optical conductivity $\sigma(\omega)$, computed with Eq.~\ref{eq:opcond} at different temperatures and normalized to its DC value. %$\sigma^{dc}$. 
%The optical conductivity exhibits a Drude-like peak centered at zero frequency, and an incoherent shoulder structure consisting of phonon sidebands plus a broad background. 
%The Drude peak is sharp at 150~K, but it broadens rapidly as the temperature increases. To compare the optical conductivity at 150 K and 300 K, %weight distribution of
%we normalize the optical conductivity by its integral at each temperature (consistent with the optical sum rule) and plot the normalized optical conductivities in Fig.~\ref{fig:optical}(b).  
%We find a significant weight transfer from the Drude peak to the incoherent shoulder as the temperature increases from 150~K to 300~K. 
%
%  TRANSPORT TIMES
%
\indent
We also extract an effective transport relaxation time, $\tau_{\text{tr}}^{*}$, from the optical conductivity through the extended Drude analysis of Ref.~\cite{Deng2014}:
\begin{equation}
\tau_{\text{tr}}^{*} = -\frac{2}{\pi\sigma^{dc}}\int_{0}^{\infty} \frac{1}{\omega^{\prime}} 
    \frac{\partial \sigma(\omega^{\prime})}{\partial \omega^{\prime}} d\omega^{\prime}.
\label{eq:tau}
\end{equation}
Figure~\ref{fig:optical}(c) shows the inverse of $\tau_{\text{tr}}^{*}$, namely the effective scattering rate characterizing the width of the Drude peak. %, is shown in Fig.~\ref{fig:optical}(c). 
%
%Note that $1/\tau_{\text{tr}}^{*}$ correspond to the QP scattering rate if transport is described within the relaxation time approximation of BTE. 
We find that this effective scattering rate increases rapidly with temperature, 
reaching values much greater than the QP scattering rate extracted from the QP peak of the spectral function in Fig.~\ref{fig:specfunc}(a). 
Due to the uncertainty principle, in a semiclassical transport regime the scattering rate cannot exceed the so-called Planckian limit of $k_{B}T$~\cite{Hartnoll2015}. We find that the effective scattering rate in SrTiO$_{3}$ exceeds the Planckian limit $k_{B}T$ above 250~K, highlighting the beyond-quasiparticle nature of charge transport in SrTiO$_{3}$ near room temperature. The breaking of the Planckian limit in SrTiO$_{3}$ at room temperature is consistent with very recent results obtained by Mishchenko~\textit{et al.} using a model Hamiltonian~\cite{Mishchenko2019}.\\
\indent
Finally, one expects the incoherent contributions to transport to be significant at temperatures where the scattering rate exceeds the Planckian limit, namely above $\sim$250~K in our calculations. 
We quantity the incoherent contribution to transport by defining the conductivity ratio $\sigma^{dc}_{\rm inc} / \sigma^{dc}$, where the incoherent contribution to the conductivity $\sigma^{dc}_{\rm inc}$ is obtained by integrating the TDF at energies greater than all QP peaks ($\omega > 32$ meV in our calculations). 
Figure~\ref{fig:specfunc}(d) shows that above 250~K the beyond-QP incoherent contributions amount to up to $\sim$40\% of the total DC conductivity, confirming that the Planckian limit breaking is associated with a fully quantum mechanical transport regime beyond the QP scattering paradigm. While the Planckian limit breaking has been typically associated with strange metals and other strongly correlated phases of matter~\cite{Emery1995,Gunnarsson2003,Bruin2013}, 
our results highlight that strong $e$-ph interactions can also lead to this quantum mechanical transport regime.
\section{Conclusion}
\vspace{-10pt}
\indent
In summary, we developed a broadly applicable approach for computing charge transport in the large polaron regime in materials with intermediate $e$-ph coupling strength. 
Our calculations on SrTiO$_{3}$ 
%can accurately predict the experimental mobility near room temperature, 
unveil a transition from band-like transport of strongly weight-renormalized QPs at low temperature to an incoherent transport regime beyond the QP picture near room temperature. 
%Our results revealed the importance of polaron effects on transport in bulk SrTiO$_{3}$, and also suggest the relevance of polaron in SrTiO$_{3}$ related structure, such as high-$T_{c}$ superconductivity in FeSe/SrTiO$_{3}$ and transport properties of LaAlO$_{3}$/SrTiO$_{3}$ heterostructure~\cite{Cancellieri2016}.
Our approach can shed new light on
%provide an approach to investigate electronic and transport properties in 
 broad classes of materials with polaron effects, ranging from perovskites~\cite{Miyata2017} and transition metal oxides~\cite{Moser2013,Cancellieri2016} to high-$T_{c}$ superconductors~\cite{Lanzara2001,Legros2019}.
%
%
%%%%%     METHODS     %%%%
%
%
\begin{acknowledgments}
\vspace{-10pt}
\noindent
J.-J.Z. has benefited from discussion with N.-E. Lee. This work was supported by the Joint Center for Artificial Photosynthesis, a DOE Energy Innovation Hub, supported through the Office of Science of the U.S. Department of Energy under Award No.~DE-SC0004993. 
M.B. acknowledges support by the National Science Foundation under Grant No.~ACI-1642443, which provided for code development, and Grant No.~CAREER-1750613, which provided for theory and method development. 
This work was partially supported by the Air Force Office of Scientific Research through the Young Investigator Program, Grant FA9550-18-1-0280.
This research used resources of the National Energy Research Scientific Computing Center, a DOE Office of Science User Facility supported by the Office of Science of the U.S. Department of Energy under Contract No. DE-AC02-05CH11231. \\%
\end{acknowledgments}
\appendix
%  SPECTRAL FUNCTIONS
\section{Finite temperature retarded cumulant approach}
\vspace{-10pt}
\label{app:cum}
%
%\newline
%\noindent
%{\bf Electron spectral function.} 
\indent
We implement and employ a retarded cumulant approach~\cite{Kas2014, Kas2017} to compute the electron spectral function at finite temperature. 
The retarded cumulant approach [see Eq.~(\ref{eq:cum_exp}) and (\ref{eq:cumulant})] was recently employed to study the electron spectral function near the band edge in insulators at zero temperature~\cite{Nery2018}. 
However, the separation scheme used in Ref.~\cite{Nery2018} to compute the cumulant and the spectral function is not well behaved at finite temperature~\cite{Kas2014}, where $\beta_{n\bm{k}}(0) > 0$. 
%
% \indent
% OUR FINITE T APPROACH
%Here, we develop a finite-temperature retarded cumulant approach. 
Here, we develop a new scheme to compute the retarded Green's function at finite temperature. 
We rewrite Eq.~(\ref{eq:cumulant}) as
\begin{equation}
C_{n\bm{k}}\left(t\right) = P\int_{-\infty}^{\infty}d \omega \frac{\tilde{\beta}_{n\bm{k}}  \left(\omega \right)}{\omega^{2}}\left(e^{-i\omega t}-1\right) - i t \Sigma_{n\bm{k}}(\varepsilon_{n\bm{k}})\,\,,
 \label{eqm:cumulant2}
\end{equation}
where $\tilde{\beta}_{n\bm{k}}  \left(\omega \right) \equiv  \beta_{n\bm{k}}(\omega) - \beta_{n\bm{k}}(0)$ and $P$ denotes the Cauchy principal value of the integral. 
The following two relations are used to derive Eq.~(\ref{eqm:cumulant2})$\,$:
\begin{equation}
\begin{split}
\int_{-\infty}^{\infty}\frac{\left(e^{-i\omega t}+i\omega t-1\right)}{\omega^{2}} d\omega  =  -\pi t  \\ 
P\int_{-\infty}^{\infty}\frac{\tilde{\beta}_{n\bm{k}} (\omega)}{\omega}d\omega =  -\mathrm{Re}\Sigma_{n\bm{k}}\left(\varepsilon_{n\bm{k}}\right). \nonumber 
\end{split}
\end{equation}
\indent
We evaluate $\tilde{\beta}_{n\bm{k}}  \left(\omega \right)$ on a discrete frequency grid, $\omega_{l}=l\Delta \omega$ with $l$ an integer.  
We label the first term in Eq.~(\ref{eqm:cumulant2}) as $C_{n\bm{k}}^{s}\left(t\right)$ and compute it from $\tilde{\beta}_{n\bm{k}}  \left(\omega_{l}\right)$, 
 \begin{equation}
C_{n\bm{k}}^{s}\left(t\right)=\sum_{\omega_{l}\neq 0}\frac{\tilde{\beta}_{n\bm{k}} (\omega_{l})}{\omega_{l}^{2}}\left(e^{-i\omega_{l}t}-1\right)\Delta\omega\,\,,
 \label{eqm:cumulant3}
 \end{equation}
Substituting Eq.~(\ref{eqm:cumulant2}) and~(\ref{eqm:cumulant3}) into Eq.~(\ref{eq:cum_exp}), the retarded Green's function becomes 
\begin{equation}
G^{R}_{n\bm{k}}(t) = -i\theta(t) e^{-i t (\varepsilon_{n\bm{k}} + \Sigma_{n\bm{k}}(\varepsilon_{n\bm{k}}))}e^{C^{s}_{n\bm{k}}\left(t\right)}.
\label{eqm:cum_exp2}
\end{equation}
% FOURIER TRANSFORMS
We then evaluate the spectral function by Fourier transforming to the frequency domain, and obtain
\begin{equation}
A_{n\bm{k}}(\omega) =  A^{0}_{n\bm{k}}(\omega) * A^{s}_{n\bm{k}}(\omega) / 2\pi\,\,,
\label{eqm:cum_exp3}
\end{equation}
where $*$ denotes a convolution operation; the two quantities entering this formula are
\begin{equation}
A^{0}_{n\bm{k}}(\omega) =
\frac{-\mathrm{Im} \Sigma(\varepsilon_{n\bm{k}}) / \pi}
{(\omega - \varepsilon_{n\bm{k}} - \mathrm{Re}\Sigma(\varepsilon_{n\bm{k}}))^{2} + {\mathrm{Im} \Sigma(\varepsilon_{n\bm{k}})}^{2}}
\end{equation}
and $A^{s}_{n\bm{k}}(\omega)$, which is the Fourier transform of $e^{C^{s}_{n\bm{k}}\left(t\right)}$. \\
\indent
Note that $C^{s}_{n\bm{k}}\left(t\right)$ defined in Eq.~(\ref{eqm:cumulant3}) is periodic with a period of $\mathcal{T} = 2\pi/\Delta \omega$; we compute $e^{C^{s}_{n\bm{k}}\left(t\right)}$ on a discrete time grid $t_{j} = j\Delta t$ with $j \in [-N, N]$ and $\Delta t = \mathcal{T}/(2N+1)$. 
It is seen from Eq.~(\ref{eqm:cumulant3}) that $C_{n\bm{k}}^{s}\left(-t_{j}\right) = C_{n\bm{k}}^{s}\left(t_{j}\right)^{*}$ and $C_{n\bm{k}}^{s}\left(0\right)=0$, which allows us to compute $C_{n\bm{k}}^{s}\left(t_{j}\right)$ only for $j \in [1, N]$ and obtain $e^{C^{s}_{n\bm{k}}\left(t_{j}\right)}$ on the full time grid. 
We then obtain $A^{s}_{n\bm{k}}(\omega)$ on a frequency grid with the same step $\Delta \omega$ as $\tilde{\beta}_{n\bm{k}} \left(\omega \right)$ via a discrete Fourier transform of $e^{C^{s}_{n\bm{k}}\left(t_{j}\right)}$. The value of $A^{s}_{n\bm{k}}(\omega)$ is guaranteed  to be real. 
We have carefully checked the convergence of $A^{s}_{n\bm{k}}(\omega)$ with respect to the range and step size of both the time and frequency grids. 
Specifically, we employed $\omega$ values ranging from $-0.8$ to $0.8$~eV with a small step $\Delta \omega$ of 0.05~meV and $N=32000$ for the time grid in our calculations.\\
%
%
% COMPUTATIONAL DETAILS
%
\section{Computational details}
\vspace{-10pt}
\label{app:detail}
%
%
%\noindent
%{\bf First-principles calculations.}
\indent
We carry out \textit{ab initio} calculations on cubic SrTiO$_{3}$ with a lattice parameter of 3.9 \AA. The ground-state electronic structure is computed within the Perdew-Burke-Ernzerhof generalized gradient approximation~\cite{Perdew2008} of density functional theory (DFT) using the {\sc Quantum ESPRESSO} code~\cite{Giannozzi2009}. We employ fully-relativistic norm-conserving pseudopotentials, which include spin-orbit coupling (SOC)~\cite{Hamann2013, Setten2018}, together with a plane-wave basis set with a kinetic energy cutoff of 85 Ry. The anharmonic lattice-dynamical properties of cubic SrTiO$_{3}$ are computed using the temperature-dependent effective potential (TDEP) method~\cite{Hellman2011} with 4~$\times$~4~$\times$~4 supercells, 
together with a scheme we recently developed~\cite{Zhou2018} to accurately include the long-range dipole-dipole contributions in the interatomic force constants.\\
%
% E-PH INTERACTIONS
%
%\newline
%\noindent
%{\bf Electron-phonon interactions.}
\indent
We use our in-house developed {\sc perturbo} code~\footnote{The code employed in this work will be released in the future at http://perturbo.caltech.edu} to compute the lowest-order $e$-ph self-energy at temperature $T$ for the Bloch state $\left|n\bm{k}\right\rangle$ with band $n$ and crystal momentum $\bm{k}$,
\begin{align}
& \Sigma_{n\bm{k}}(\omega, T) =\sum_{m,\nu\bm{q}}\left|g_{mn,\nu}(\bm{k},\bm{q})\right|^{2}  \times  \nonumber \\
 &  \left[\frac{N_{\nu\bm{q}}+f_{m\bm{k}+\bm{q}}}{\omega-\varepsilon_{m\bm{k}+\bm{q}}  +\omega_{\nu\bm{q}}+i\eta}
 + \frac{N_{\nu\bm{q}}+1-f_{m\bm{k}+\bm{q}}}{\omega-\varepsilon_{m\bm{k}+\bm{q}}-\omega_{\nu\bm{q}}+i\eta} \right] \,\,,
 \label{eqm:ep_selfe}
\end{align}
where $\varepsilon_{n\bm{k}}$ is the DFT band electron energy, $\omega_{\nu\bm{q}}$ is the energy of a phonon with branch index $\nu$ and wave vector $\bm{q}$, $f_{n\bm{k}}$ and $N_{\nu\bm{q}}$ are, respectively, 
the Fermi and Bose occupation numbers evaluated at temperature $T$, and $\eta$ is a small broadening.
The key quantities are the $e$-ph coupling matrix elements, $g_{mn,\nu}(\bm{k},\bm{q})$, defined as 
\begin{equation}
g_{mn,\nu}\left(\bm{k},\bm{q}\right)=\sqrt{\frac{\hbar}{2\omega_{\nu\bm{q}}}}\sum_{\kappa\alpha}\frac{\bm{e}_{\nu\bm{q}}^{\kappa\alpha}}{\sqrt{M_{\kappa}}}\left\langle m{\bm{k}+\bm{q}}\left|\partial_{\bm{q}\kappa\alpha}V\right|n\bm{k}\right\rangle,
\label{eqm:eph_mat}
\end{equation}
where  $\partial_{\bm{q}\kappa\alpha}V\equiv \sum_{p} e^{i\bm{q}\bm{R}_{p}} \partial_{p\kappa\alpha}V$, with $\partial_{p\kappa\alpha}V$ the variation of the Kohn-Sham potential for a unit displacement of the atom $\kappa$ (with mass $M_{\kappa}$ and in the unit cell at $\bm{R}_{p}$) in the $\alpha$-direction, and $\bm{e}_{\nu\bm{q}}$ is the phonon displacement eigenvector. 
We obtain the $e$-ph matrix elements with the method described in our recent work~\cite{Zhou2018}. Briefly, we first compute the electronic wave functions $\left|n\bm{k}\right\rangle$ on an 8~$\times$~8~$\times$~8 $\bm{k}$-point grid using DFT, construct Wannier functions for the Ti-$t_{2g}$ orbitals from the three lowest conduction bands using the {\sc Wannier90} code~\cite{Mostofi2014}, compute $\partial_{\bm{q}\kappa\alpha}V$ on an 8~$\times$~8~$\times$~8 $\bm{q}$-point grid with density functional perturbation theory~\cite{Baroni2001}, and compute anharmonic phonon energies $\omega_{\nu\bm{q}}(T)$ and eigenvectors $\bm{e}_{\nu\bm{q}}(T)$ with TDEP. 
Using these ingredients, we first evaluate the $e$-ph matrix elements in Eq.~(\ref{eqm:eph_mat}) on coarse grids. Wannier interpolation together with a long-range $e$-ph correction for the polar modes~\cite{Giustino2007,Sjakste2015,Verdi2015} is then employed to interpolate $g_{mn,\nu}(\bm{k}\bm{q})$ for $\bm{k}$- and $\bm{q}$-points on very fine Brillouin zone grids, which are needed to converge the $e$-ph self-energy in Eq.~(\ref{eqm:ep_selfe}). 
An approach we recently developed~\cite{Zhou2016} is employed to converge the imaginary part of the self-energy, $\mathrm{Im}\Sigma_{n\bm{k}}(\omega)$, which is computed off-shell on a fine energy $\omega$ grid, 
and the real part of the self-energy, $\mathrm{Re}\Sigma_{n\bm{k}}$, which is evaluated on-shell at the electron energy $\varepsilon_{n\bm{k}}$. 
The spectral functions $A_{n\bm{k}}(\omega)$ are then obtained using the finite temperature retarded cumulant approach described in Appendix~\ref{app:cum}, using $\mathrm{Im}\Sigma_{n\bm{k}}(\omega)$ and $\mathrm{Re}\Sigma_{n\bm{k}}(\varepsilon_{n\bm{k}})$ as input.  \\
\indent
We compute spectral functions $A_{n\bm{k}}(\omega)$ on fine $\bm{k}$-point grids with up to 100$^{3}$ points in the Brillouin zone, which are then employed in the conductivity and mobility calculations using Eq.~(\ref{eq:opcond}) and~(\ref{eq:nc}). 
Note that the spectral functions depend on both the temperature $T$ and chemical potential $\mu$. For each temperature and carrier concentration $n_{c}$ considered in our calculations, we determine self-consistently the chemical potential $\mu(T, n_{c})$ using Eq.~(\ref{eq:nc}); this means that the spectral functions computed using the chemical potential $\mu(T, n_{c})$ give a consistent carrier concentration $n_{c}$.  \\
\begin{figure}[!t]
\includegraphics[width=0.8\columnwidth]{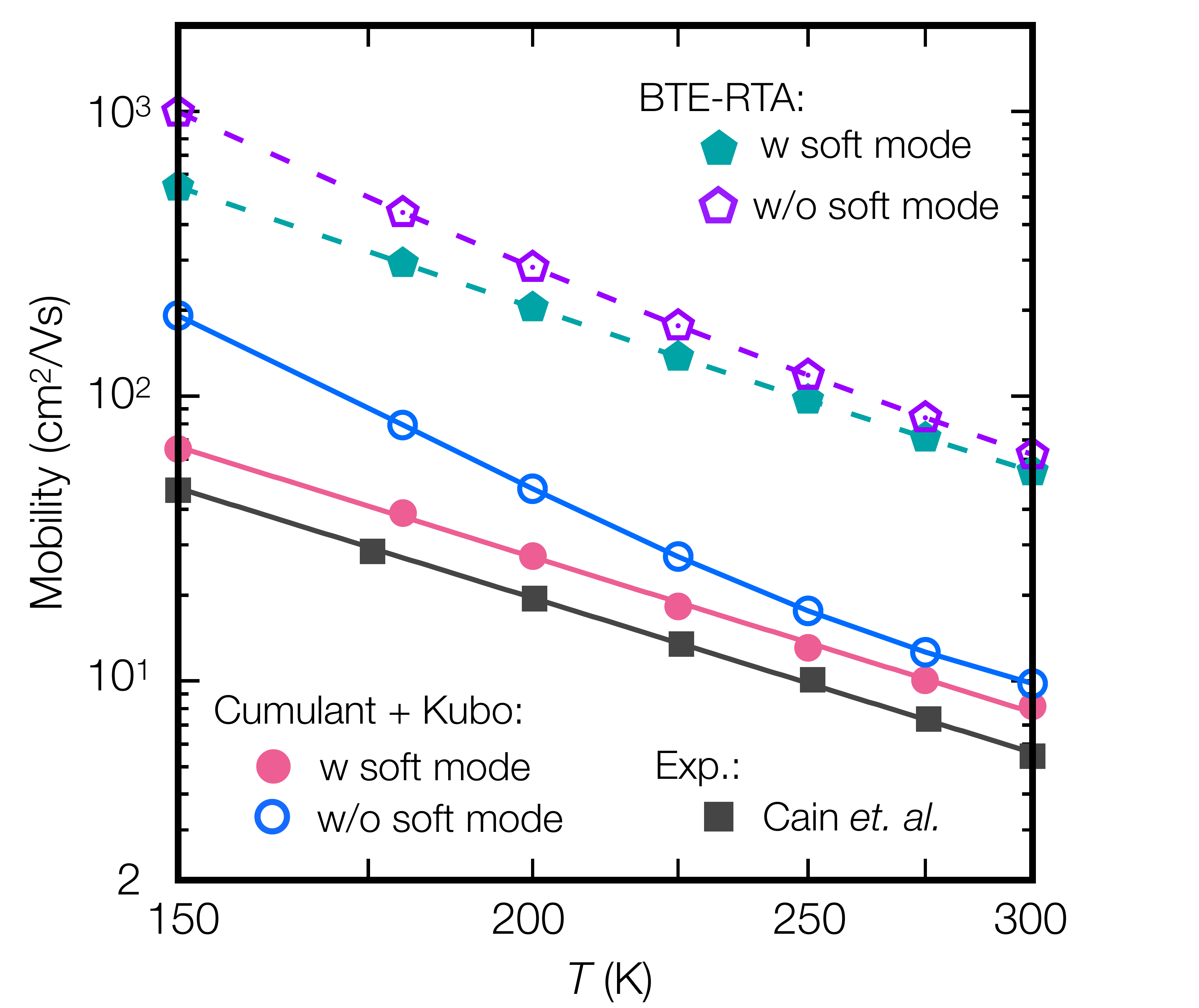}
\caption{Electron mobility in SrTiO$_{3}$ computed using the BTE within the RTA and the cumulant approach plus the Kubo formula, compared with experimental data~\cite{Cain2013}. For both the BTE-RTA and the cumulant approach, the electron mobility obtained by excluding the contribution from the ferroelectric soft phonon mode is also shown. 
} \label{fig:soft}
\end{figure}
\vspace{-10pt}
\section{Soft mode contribution to transport}
\vspace{-10pt}
\label{app:softmode}
To highlight the contribution from the soft modes to charge transport in SrTiO$_{3}$, we compare the electron mobility computed with and without the soft mode contribution, as shown in Fig.~\ref{fig:soft}. Our recent calculations using the BTE show that the ferroelectric soft mode plays an important role in transport, especially at low temperatures below 250~K, and that including the soft mode contribution is critical to obtaining an accurate temperature dependence of the electron mobility~\cite{Zhou2018}. The results from the cumulant approach exhibit the same trend $-$ neglecting the soft mode contribution leads to a significant overestimate of the mobility below 250~K and to an inaccurate temperature dependence. Although the phonon satellites in the spectral function is primarily due to the strong coupling between the electron and LO phonons, the ferroelectric soft mode can impact the width of the QP peak, thus contributing to charge transport. 
\begin{figure}[!t]
\includegraphics[width=\columnwidth]{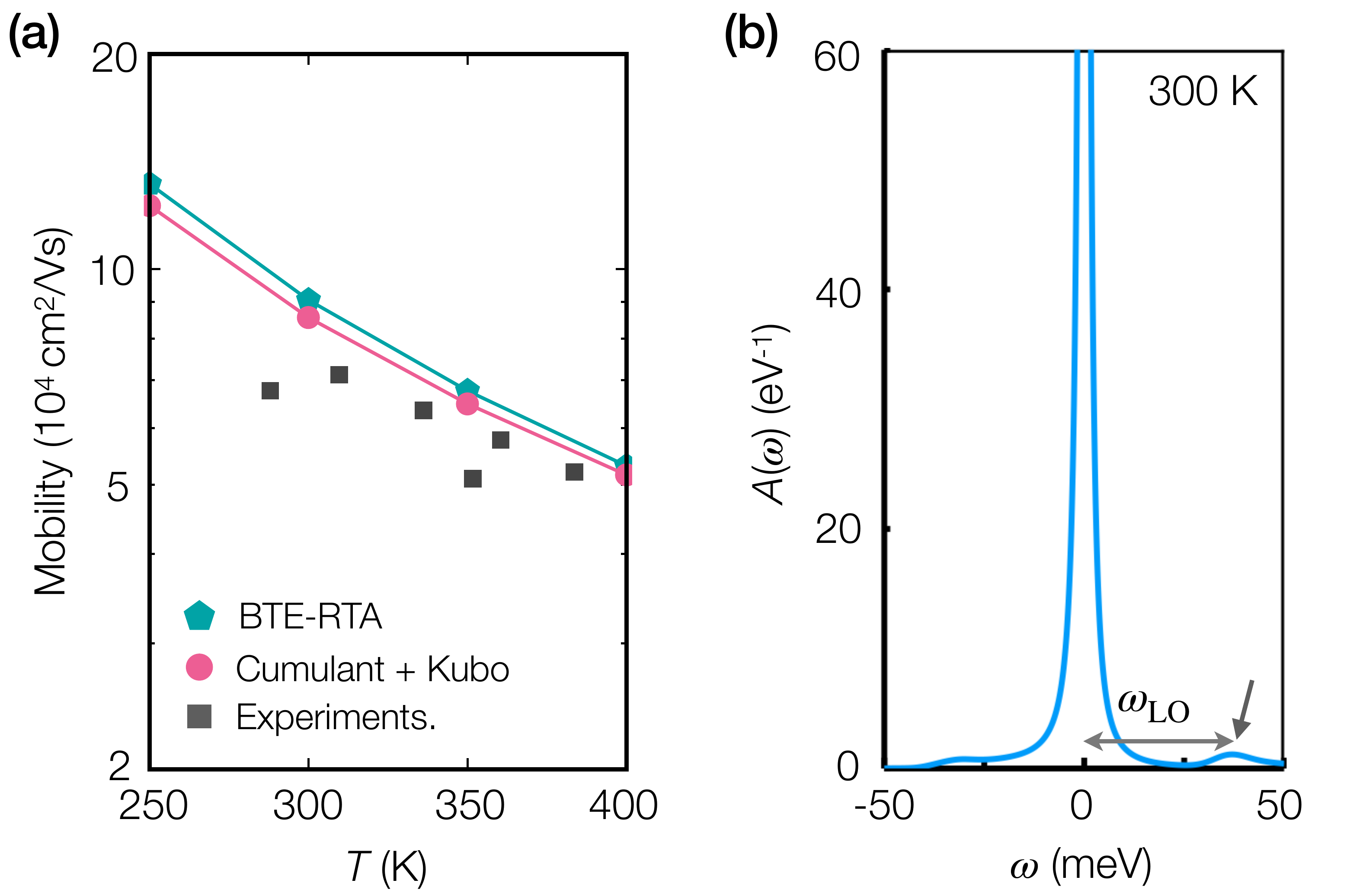}
\caption{ (a) Electron mobility in GaAs as a function of temperature, computed using the cumulant approach plus the Kubo formula (red circles), the BTE approach within the RTA (blue pentagons), and compared with experimental values taken from Refs.~\cite{Blood1972,Nichols1980}. (b) Spectral function for the electronic state at the conduction band minimum in GaAs, computed at 300~K using the cumulant approach. Note the small phonon sidebands at energy $\omega_{\rm LO} = 36$ meV above the QP peak. 
} \label{fig:gaas}
\end{figure}
\vspace{-10pt}
\section{Computed electron mobility in GaAs: BTE vs cumulant approach}
\vspace{-10pt}
\label{app:gaas}
\indent
The method we developed in this work, which uses the retarded cumulant approach plus the Kubo formula to compute charge transport, 
is general and can be applied broadly in materials with $e$-ph interactions ranging from weak to strong. 
To illustrate this point, we apply our approach to GaAs, a high mobility semiconductor in which large-polaron effects are very weak. 
We have recently computed the temperature dependent electron mobility in GaAs using the BTE approach within the RTA~\cite{Zhou2016}. With the same computational settings for the DFT and DFPT calculations as in Ref.~\cite{Zhou2016}, we compute the electron mobility at temperatures between 250$-$400~K using the cumulant approach. Figure~\ref{fig:gaas}(a) compares the electron mobility from the cumulant and BTE-RTA calculations with experimental data.
Due to the weak $e$-ph coupling, the spectral function in GaAs consists of a sharp QP peak with near-unit spectral weight [see Fig.~\ref{fig:gaas}(b)], plus a weak phonon sideband at energy $\omega_{\rm LO} = 36$ meV above the QP peak, where $\omega_{\rm LO} $ is the LO phonon energy in GaAs. 
The cumulant approach and the BTE-RTA give mobility results in close agreement with each other, as is expected in materials with weak $e$-ph coupling. 
We have thus shown that our implementation of the cumulant plus Kubo approach without current-vertex corrections reduces to the RTA solution of the BTE in the weak $e$-ph coupling limit~\cite{Economou2006,Allen2015}.\\
\indent
For linear-response theory to be consistent with the BTE, the Kubo formula with current-vertex corrections should reduce in the weak $e$-ph coupling limit to the full solution of the BTE, e.g. the solution obtained in practice with an iterative method~\cite{Ma2018}. Our calculation on GaAs in the absence of current-vertex corrections shows good agreement with the BTE-RTA, and similarly we expect that including the current-vertex corrections would give results consistent with the iterative BTE solution for GaAs. In the case of SrTiO$_{3}$, the BTE-RTA and the iterative solution of BTE produce very similar results as we have shown previously~\cite{Zhou2018}. Therefore, the current-vertex corrections are not essential for SrTiO$_{3}$ and are neglected in our calculations. \\
%\newline
\bibliography{ref-cumulant}
\end{document}